# Cancer driver gene detection in transcriptional regulatory networks using the structure analysis of weighted regulatory interactions

Mostafa Akhavan Safar [1], Babak Teimourpour [2,*], Abbas Nozari-Dalini [3]

[1,2] Department of Information Technology Engineering, School of Systems and Industrial Engineering, Tarbiat Modares University (TMU), Tehran, Iran.

Email: m.akhavansafar@modares.ac.ir

[*] **Corresponding author**. Assistant Professor, Room No. 210, 2nd Floor, Information Technology Engineering group, School of Systems and Industrial Engineering, Tarbiat Modares University (TMU) Chamran/Al-e-Ahmad Highways Intersection, Tehran, P.O. Box 14115-111, Iran.
Tel: +98(21) 8288 3987
E-mail address: b.teimourpour@modares.ac.ir

[3] Department of Computer Science, School of Mathematics, Statistics, and Computer Science, University of Tehran, Tehran, Iran.
E-mail address: nowzari@ut.ac.ir

**Abstract:**

Identification of genes that initiate cell anomalies and cause cancer in humans is among the important fields in the oncology researches. The mutation and development of anomalies in these genes are then transferred to other genes in the cell and therefore disrupt the normal functionality of the cell. These genes are known as cancer driver genes (CDGs). Various methods have been proposed for predicting CDGs, most of which based on genomic data and based on computational methods. Therefore, some researchers have developed novel bioinformatics approaches. In this study, we propose an algorithm, which is able to calculate the effectiveness and strength of each gene and rank them by using the gene regulatory networks and the stochastic analysis of regulatory linking structures between genes. To do so, firstly we constructed the regulatory network using gene expression data and the list of regulatory interactions. Then, using biological and topological features of the network, we weighted the regulatory interactions. After that, the obtained regulatory interactions weight was used in interaction structure analysis process. Interaction analysis was achieved using two separate Markov chains on the bipartite graph obtained from the main graph of the gene network. To do so, the stochastic approach for link-structure analysis has been implemented. The proposed algorithm categorizes higher-ranked genes as driver genes. The efficiency of the proposed algorithm, regarding the F-measure value and number of identified driver genes, was compared with 23 other computational and network-based methods. We used four validated CDG databases to correctitude the results. Results indicate that the proposed algorithm outperforms other methods and is capable of identifying a significant number of driver genes that were not identified using other methods.

**Keywords**: driver genes, cancer, stochastic approach for link-structure analysis, gene regulatory network.

## 1. Introduction

Many kinds of researches have been conducted regarding the detection of cancer-causing genes [1-3]. According to various conducted researches on identifying ways to fight cancer, many researchers utilize a high-technology process called "genome sequencing" that simulates genetic mutation to investigate genes, which are linking to the growth of tumor cells. Therefore, due to the limitations of sequencing methods, some researchers have developed novel bioinformatics methods, each of which helps to detect cancer mutations. According to the results of the previous researches, new methods with better results are still needed. The proposed methods for the identification of these genes are categorized as computational and network-based methods. The majority of proposed methods, which identify driver genes based on the number of mutations and investigation of the genomic data, are categorized as computational methods. Some of the proposed computational methods include Simon [4], OncodriveFM [5], ActiveDriver [6], DrGap [7], iPAC [8], MutSigCV [9], OncodriveCLUST [10], e-Driver [11], MSEA [12], ExInAtor [13] and rDriver [14], among which Simon, OncodriveFM,

ActiveDriver, DrGap, MutSigCV, ExInAtor and rDriver, Dendrix [15] and DriverML identify cancer driver genes by examining the level of background mutation rate. Another set of computational methods such as iPAC, OncodriveCLUST, e-Driver, and MSEA identify driver genes by examining the mutation domains, which are usually located on the functional domains of the protein or significant areas of the protein structure. These methods mostly involve complicated calculations and high expenses and, in terms of efficiency, have a low F-measure.Therefore, the use of network-based methods and network structure to find CDGs was proposed, which was able to eliminate some existent problems of the computational procedures. These methods can be divided in two ways. Methods which are a combination of computational and network-based methods such as NetBox, DriverNet, MDPFinder, MEMo, CoMDP, DawnRank, SCS, iMaxDriver-N and iMaxDriver-W, among which NetBox, Dendrix, DriverNet, MDPFinder, MEMo, DawnRank and SCS have used available knowledge of the protein interaction networks, molecular networks and signaling pathways, along with the concept of mutation. Other methods, such as iMaxDriver-N and iMaxDriver-W have identified driver genes using only network-based methods, without using the concept of mutation and pathway. These methods have shown a relatively appropriate level of performance, without using genomic and mutations data, but still, involve some limitations. For example, the two methods of iMaxDriver-N and iMaxDriver-W are sluggish and time-consuming in terms of computation. Furthermore, due to the nature of algorithms, which are based on diffusion maximization, only driver genes of the TF[1] type will be identified; however, some of the driver genes are the mRNA type, which cannot be recognized through these methods. Considering the significance of the subject, research in this area continues to optimize the performance of the proposed algorithms. In this study, we suggest a network-based approach by using the concept of random Markov chains and the structure of gene regulation networks without the use of genomic data and mutations. This method can classify genes into two categories, CDG and normal, only by randomly analyzing the structure of regulatory interactions without using path data and the concept of mutation. The application of stochastic analysis approach for weighted regulatory interactions has not been implemented yet in gene regulatory networks to identify cancer driver genes. We compare the results of the proposed method in the present study with 23 other aforementioned computational and network-based methods, and we will reveal that the proposed method outperforms other methods in most cases.

2. **Stochastic Markov chains and link analysis of graphs**

The Markov chain is a stochastic model that can be used for describing a sequence of probable events. Within this sequence, the probability of each event depends on the state of the previous event. Markov Chain is, in fact, a mathematical system in which transitions between states take place from one state to another. The Markov chain is a random process without memory, meaning that the conditional probability distribution of the next state depends on the current state. This type of lack of memory is called the Markov characteristic. The entire World Wide Web can be considered a Markov chain in which each page is a state and the link between them indicates the probability of each transition. This theory signifies that independent of the pages from which we have started, after a long duration of browsing the web, the probability of reaching a particular page is of a constant value. With this type of modeling, it could be stated that the higher likelihood of reaching a page indicates the higher significance of that page. An algorithm, which is developed by the Google Company by using the Markov chain, is the PageRank algorithm that ranks the web pages This algorithm assigns a number to each element of a set of hyperlinks and can allocate a degree of weight to each entity in a set, given the interactions between entities. The numerical weight assigned to each element V indicates the PageRank of page V, shown as PR (V). These ranking values are implemented by applying a Markov chain on the web graph. The rank value indicates the importance of a particular page. An output link to a page is considered a numerical value for increasing its rank. The rank of a page is defined recursively and depends on the number and rank of all pages that link to it (input links). If a page is linked to many pages with high rank, that page will also increase in rank. Another algorithm, which is useful on the internet based on the Markov chain, is the HITS [16] algorithm that uses a Markov chain to calculate two ranks called hub and authority for each node. In this algorithm, some pages are operating as central or main pages (hub), such as pages that have many output links; other pages are subject-based, meaning that they possess many internal links. This algorithm assigns two scores to each page, i.e., hub and authority. According to Kleinberg's pattern [30], good hub pages point to good authorities and in return, good authorities refer to good hub pages. Another algorithm for ranking webpages is "the stochastic approach for link-structure analysis (SALSA)" . Twitter uses the modified version of this algorithm to make the following suggestions to users. This algorithm uses two independent Markov chains, which will be introduced in the next section. We used the modified version of this algorithm to find cancer driver genes in gene regulatory networks.

---

[1] Transcription factor

### 3. Gene regulatory network and interactions structure analysis approach

The stochastic approach for link-structure analysis, proposed by S. Moran and R. Lempel, is a webpage-ranking algorithm that assigns scores to the hub and authority nodes based on the number of links between them. This approach operates based on the Markov chain theory and in fact, utilizes the stochastic properties of random walks on webpages. This algorithm combines the random walk approach of Google's PageRank algorithm with the hub and authority technique of the HITS algorithm. In this approach, as depicted in Figure 1, the graph is firstly converted to a bipartite graph based on in-degree and out-degree of vertices. Then, two separate random walks are carried out in this bipartite graph. Each walk is limited on only one side of the graph, and two walks naturally start from different sides of the bipartite graph. In contrast to ordinary random walks, this approach executes transition in Markov chains by passing through two links. Analysis of both chains allows the algorithm to assign two separate scores to each node, i.e., a hub score and an authority score. Therefore, two different Markov chains related to these random walks will be investigated: A chain of visits on the authority side of the graph and a chain of visits on the hub side of the graph. A separate analysis of these chains will naturally create changes among these two concepts for each node.

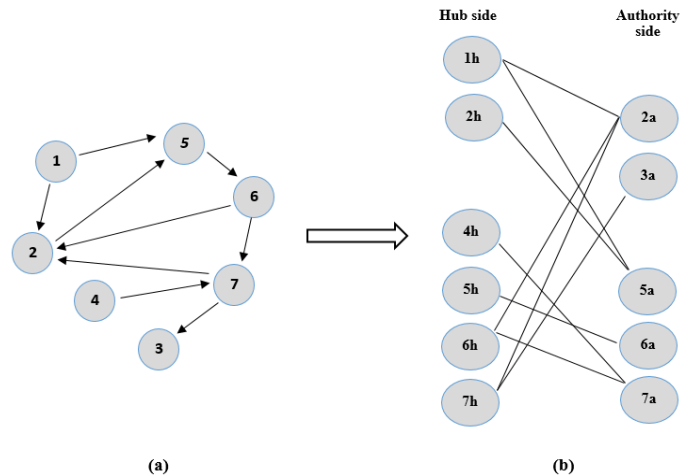

**Figure 1.** (a) Directed graph with 7 nodes and 9 edges. (b) The respective bipartite graph.

Assuming that the directed graph G = (V, E), in which V is the non-isolated vertex set and E is the sum of edges, the bipartite graph related to the graph G is shown as $G_b(V_h, V_a, E)$, and let all edges be directed from nodes in $V_h$ to nodes in $V_a$.

*The hub side of $G_b$:*
$V_h = \{c_h \mid c \in V \text{ and } out-degree(c) > 0\}$
*The authority side of $G_b$ :*
$V_a = \{c_a \mid c \in V \text{ and } in-degree(c) < 0\}$
$E = \{(c_h, c_a) \mid c_h \to c_a\}$, →: represents an edge between two vertices

Some nodes in G have in-degree and out-degree greater than zero, these nodes belong to both $V_h$ and $V_a$.

Two separate random walks are carried out in this bipartite graph. Each walk visits nodes only from one side of the graph by going through pathways, including two edges of $G_b$, at each stage. Since each edge passes through one side of $G_b$, each walk is limited to only one side of the graph, and two walks start off naturally from different sides of $G_b$. For the Markov chain of authority side with the set of states A, the transmission probability between both states of $i_a, j_a \in A$ is as follows:

$$P_A(i_a, j_a) = \sum_{\substack{k_h \to i_a \\ k_h \in G_b}} \sum_{\substack{k_h \to j_a \\ k_h \in G_b}} \frac{1}{in\_deg(i_a)} \times \frac{1}{out\_deg(k_h)} \tag{1}$$

The simulation of user effort is a random walk on the web graph. We can also compute the probability of arriving at a particular node, $j_a$, as follows:

$$P_A(go\ to\ j_a) = P_A(follow\ path\ i_a \rightarrow j_a|\ at\ i_a).P_A(i_a)$$

The basic idea is finding a stationary distribution over the graph. This is done by solving the following linear equation system: find a non-negative number, $A(j_a)$, for each vertex, $j_a$, in authority side such that:

$$A(j_a) = \sum_{(k_h,i_a)\in G_b} \sum_{(k_h,j_a)\in G_b} \frac{A(i_a)}{out\_deg(k_h)in\_deg(i_a)} \qquad (2)$$

Similarly, for the Markov chain of hub side:

$$P_H(i_h,j_h) = \sum_{\substack{(i_h\rightarrow k_a) \\ k_a\in G_b}} \sum_{\substack{(j_h\rightarrow k_a) \\ k_a\in G_b}} \frac{1}{out\_deg(i_h)} \times \frac{1}{in\_deg(k_a)} \qquad (3)$$

Therefore, two independent algorithms based on separate Markov chains are used to calculate hub and authority scores.

If the network has weighted edges, we can modify the links' analytical stochastic equations as follows. The same as before, for the Markov chain of authority side with the set of states A, the transmission probability between both states of $i_a, j_a \in A$ is as follows:

$$P_A(i_a,j_a) = \sum_{\substack{k_h\rightarrow i_a \\ k_h\in H}} \sum_{\substack{k_h\rightarrow j_a \\ k_h\in H}} \frac{w_{k_h\rightarrow i_a}}{w_{in\_deg}(i_a)} \times \frac{w_{k_h\rightarrow j_a}}{w_{out\_deg}(k_h)} \qquad (4)$$

Similarly, for the Markov chain of hub side with the set of states H, the transmission probability between both states of $i_h, j_h \in H$ is as follows:

$$P_H(i_h,j_h) = \sum_{\substack{i_h\rightarrow k_a \\ k_a\in A}} \sum_{\substack{j_h\rightarrow k_a \\ k_a\in A}} \frac{w_{i_h\rightarrow k_a}}{w_{out\_deg}(i_h)} \times \frac{w_{j_h\rightarrow k_a}}{w_{in\_deg}(k_a)} \qquad (5)$$

In which:

$$\forall j \in H: \ w_{out\_deg}(j) = \sum_{\{\forall\ i\in A\ |\ j\rightarrow i\}} w_{j\rightarrow i} \qquad (6)$$

The gene regulatory network (GRN) includes a set of genes that interact with other fragments within the cell through RNA and their expression products, i.e., proteins. Therefore, it specifies the transformation rate in the network that shows which gene must be transcribed to mRNA and the amount of transcription. Generally, each molecule of the mRNA is translated into a specific protein or a set of proteins. Therefore, available genes in the gene regulatory network lead to the decrease or increase of each other's expression, and as a result, protein production, through the positive or negative impacts that they have on each other. These interactions are such that cell performance maintains a normal state. If due to irregularities and anomalies, one of the genes produces proteins outside of the regulatory conditions, it will cause this anomaly to transfer to other genes through protein-protein interactions, leading to disruption of the regulatory network. This disruption of cell performance leads to cancer. Identifying these genes, which generally initiate anomalies and cancers and are known as driver genes, is of great importance.

In this study, we will use the gene regulatory network for the stochastic investigation and analysis of weighted interactions. In this network, Tfs and mRNAs are considered as nodes, and the directed regulatory interactions between them are assumed as edges of the network. An example of the studied network is illustrated in Figure 2.

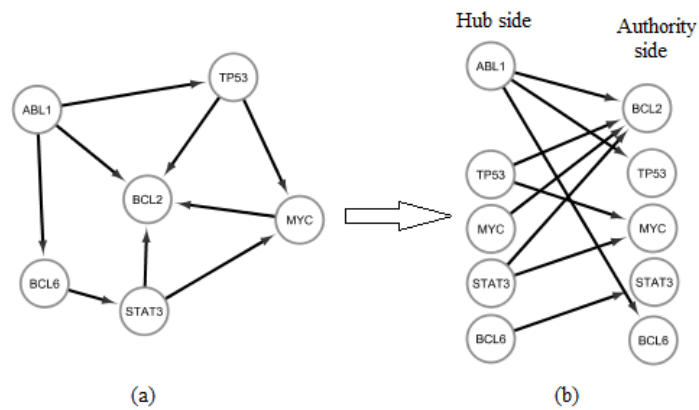

**Figure 2.** (a) Gene regulatory graph with 6 genes and 9 regulatory interactions. (b) Bipartite gene regulatory graph.

4. **Research methodology**

The goal is to calculate the significance of each gene in the network, by using the stochastic analysis of weighted regulatory interactions. We assumed the genes that have more impressibility (i.e., the power of their input regulatory interactions is higher), and more influencing power (i.e., the power of their output regulatory interactions is higher), are more likely to produce anomalies by altering their performance or disrupt the network performance due to the higher permeability from the other genes. Therefore, through the stochastic analysis of the structure of regulatory interactions, genes can be ranked based on their input and output regulatory strength. The application of the stochastic analysis approach for weighted regulatory interactions to identify influential genes has not been implemented in gene networks yet. Several diffusion-based algorithms, such as iMaxDriver, have been suggested so far, which have some weaknesses. In these methods, only the gene output flow is considered as the diffusion rate of a node; while, the input flow of each gene is not considered. Therefore, in these algorithms, driver genes are identified only among genes with above-zero output, which is one of the bugs of these methods. Since in a transcription regulatory network, source genes are mostly of the TF type, and target genes are often of the mRNA type, these methods are only successful in identifying driver genes of the TF type, and mRNA driver genes are usually overlooked. However, cancer-causing genes are of both mRNA and TF types. By implementing the stochastic analysis approach of regulatory interactions, the weaknesses of previous influence-based methods are expected to be eliminated, because the proposed method considers a gene's both input and output interaction strengths. However, other algorithms based on the Markov chain, such as the PageRank, only examine the nodes' output and accordingly identify significant nodes. In addition to using the stochastic analysis approach of regulatory interaction for gene networks, we also considered the interactions weight in calculations.

### 4.1. Construction the gene regulatory networks:

We used the list of regulatory interactions to construct regulatory networks. The list of regulatory interactions available at the TRRUST v2 [18] database. This database contains 8427 TF-TF and TF-mRNA regulatory interactions. Furthermore, it contains 2862 genes including TF (795) and mRNA (2067). For weighting the nodes and edges of the gene regulatory networks, cancer gene expression data were required. To do so, the gene expression values related to breast invasive carcinoma (BRCA), was downloaded from the GEO (Gene Expression Omnibus) database. The expression data has been reported separately for normal and cancerous tissues of patients in this database. Before using these files, they needed to be preprocessed. To do so, firstly, synonymous genes were isolated and iterative gene values were averaged. Therefore, a gene regulatory network was built using the above data, such that for breast cancer network, the source and target genes were searched in the gene expression data related to the same type of cancer. If both the source and the target possessed gene expression data, the interaction was kept, and if not, that interaction was removed from the final list.

### 4.2 Weighting the network

To calculate the weight of interactions, we considered both of the biological and topological features of the network. Thus, two weights were obtained for each interaction. We used gene expression and Pearson correlation coefficient to calculate biological weights as follows:

$$BW(gene_i, gene_j) = \frac{\sum_{k=1}^{n}(g(gene_i,k) - \bar{g}(gene_i)) \cdot (g(gene_j,k) - \bar{g}(gene_j))}{(n-1)S_{gene_i}S_{gene_j}} \quad (7)$$

In the above equation, $g(gene_i, k)$ represents the k$^{th}$ expression value of the $gene_i$. Also, $S_{gene_i}$ represent the standard deviation of the expression values for $gene_i$ and is calculated as follows:
In doing so, the weight of regulatory interactions was calculated based on the expression values for the source and target genes of each edge.

To calculate edge weight based on network topology, same as [17], the principle of equilibrium flow in the network theory was used. To do so, firstly, a weight was assigned to each gene in the network based on the absolute value of expression difference for normal and cancerous tissues. According to this, the influence rate of one gene on the genes that it regulates (weight of target gene) equals regulatory interaction multiplied by the weight of the target node. Therefore, if $e$ is the weight vector for network nodes, and M is the adjacency matrix of the graph, then:

$$\vec{e} = M^T . \vec{e} \qquad (8)$$

Vector $\vec{e}$ has known values and $M^T$ is transpose of adjacency matrix in which we replaced elements 1 with random values as regulatory interactions weight. These random values are updated when solving the equation. For solving the matrix equation above, $M^T.\vec{e} - \vec{e} = 0$ was solved. To do so, $\|M^T.\vec{e} - \vec{e}\|$ was minimized. For example, for the gene graph shown in Figure 2, it is as follows:

$$minimize \|M^T e - e\| = \left( \sum_{i=1}^{n} (\sum_{j=1}^{n} a_{ij} e_j - e_i)^2 \right)^{1/2} \qquad (9)$$

By solving equation (13), a matrix containing the weight of the interactions is obtained. We named these weights as topological weights (TW). After calculating the topology-based weights (TW) of the network, the regulatory interaction weight between $i$ and $j$ ($IW_{gene(i) \to gene(j)}$) was calculated as follows:

$$IW_{gene(i) \to gene(j)} = BW\big(gene(i), gene(j)\big) + TW\big(gene(i), gene(j)\big) \qquad (10)$$

Now, by calculating the weight of regulatory interactions, the probability of state transition between genes in built gene regulatory networks was calculated based on the following equation:

$$P_A(gene(i), gene(j)) = \sum_{\substack{gene(k_h) \to gene(i_a) \\ gene(k) \in H}} \sum_{\substack{gene(k_h) \to gene(j_a) \\ gene(k) \in H}} \frac{IW_{gene(k_h) \to gene(i_a)}}{IW_{in\_deg}(gene(i_a))} \times \frac{IW_{gene(k_h) \to gene(j_a)}}{IW_{out\_deg}(gene(k_h))} \qquad (11)$$

$$P_H(gene(i), gene(j)) = \sum_{\substack{gene(i_h) \to gene(k_a) \\ gene(k) \in A}} \sum_{\substack{gene(j_h) \to gene(k_a) \\ gene(k) \in A}} \frac{IW_{gene(i_h) \to gene(k_a)}}{IW_{in\_deg}(gene(i_h))} \times \frac{IW_{gene(j_h) \to gene(k_a)}}{IW_{out\_deg}(gene(k_a))} \qquad (12)$$

$$IW_{in\_deg}(gene(i)) = \sum_{\{\forall k \in H \mid k \to i\}} IW_{gene(k) \to gene(i)} \qquad (13)$$

To calculate the scores for each gene, first, the graph for cancer network was converted into bipartite graphs, and the analysis of weighted regulatory interactions was performed by running the algorithm.

Then, to calculate the final score for each gene, the hub score and authority score were combined as follows.

$$Score(gene_i) = \beta \times H(gene_i) + (1 - \beta) \times A(gene_i) \qquad (14)$$

After calculating the score of each gene and sort them in descending order, we classified them into driver gene and normal based on a threshold value. To determine the precise value of the $\beta$ parameter, the area under the ROC curve was used. The area under curve was calculated for different values of $\beta=\{0.1,0.2,0.3,0.4,0.5,0.6,0.7,0.8,0.9,1.0\}$ on hub and authority scores obtained from the running of the algorithm. As illustrated in Figure 4, the best results are when $\beta=0.9$.

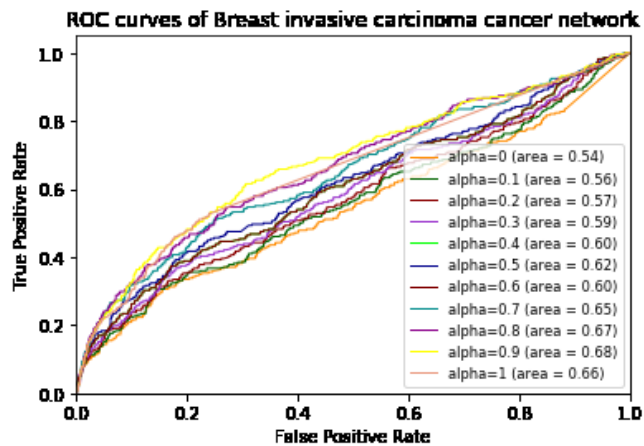

**Figure 4.** ROC graph of algorithm output for different $\beta$ values

## 5. Evaluation method

The proposed algorithm was compared with 23 other computational and network-based methods available for CDG prediction. The standard driver gene databases were required for the validation of predicted genes. We used various standard and verified data sets (Table 2):

1- The set of cancer genes introduced in the Cancer Gene Census (CGC) for Breast invasive carcinoma (TCGA-BRCA), which is available via https://cancer.sanger.ac.uk/census. In this standard database for breast invasive carcinoma, colon adenocarcinoma, and Lung squamous cell carcinoma, 572, 572, and 566 driver genes have been reported, respectively.
2- Validated Mut-driver genes, introduced by Vogelstein et al., in this dataset, were obtained from the respective article, and 125 driver genes have been reported.

Also, the F-measure criteria, which is a prevalent criterion for evaluating classification problems that consider both Recall and Precision criteria, was used to examine the performance of the proposed algorithm and compare it with other methods. The Recall criterion shows the ratio amount of the number of genes correctly detected as driver genes, to the total number of driver genes. Precision indicates prediction accuracy and is equivalent to the number of driver genes identified as driver genes. Precision evaluates the ratio amount of the number of "accurate predictions" for samples of a specific class to the "total number of predictions" for samples of the same specific class

$$\text{Recall} = \frac{TP}{TP + FN} \quad (15)$$

$$\text{Precission} = \frac{TP}{TP + FP} \quad (16)$$

$$F - \text{measure} = 2 \times \frac{\text{Precision} \times \text{Recall}}{\text{Precision} + \text{Recall}} \quad (17)$$

**Table 2.** Database used for evaluation of the proposed method and other methods.

|  | CGC (Cancer Gene Census) | Mut-driver (Vogelstein et al) |
|---|---|---|
| **Breast invasive carcinoma** | 572 | 125 |

## 6. Results

By using the regulatory interaction data and gene expression data, and after the required preprocessing, the human gene regulatory network was built. Afterward, by using gene expression data and network structure, regulatory interactions were weighted separately for each network. Afterward, the proposed algorithm was applied separately on each network to calculate the rank of each gene. The results of the proposed algorithm were obtained as a list of driver genes arranged in a descending order based on their rank. Then, by interpreting the results based on a threshold value, genes were categorized into two groups of normal and driver. The proposed algorithm and 23 other methods were evaluated by using the mentioned evaluation criteria in the previous section. Results related to breast carcinoma using the CGC database are depicted in Figure 5. As observed, the proposed algorithm has the highest F-measure value compared to other network-based and computational methods and is the most efficient. It also has been able to predict 183 genes as drivers, which after iPac [8] holds the highest record for the number of detected driver genes. Moreover, it has obtained the first rank in comparison to the network-based methods.

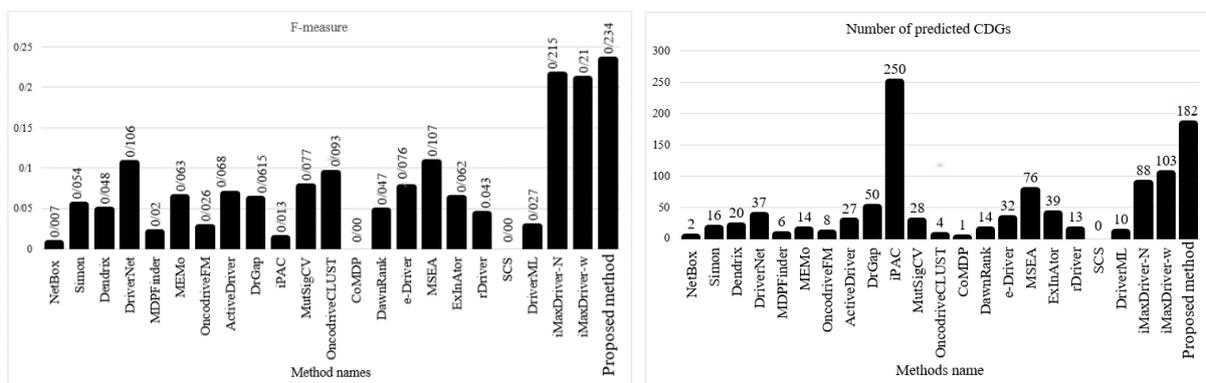

**Figure 5.** Results of the proposed algorithm and 23 other methods for breast invasive carcinoma (using CGC datasets).

Furthermore, we have used the Venn diagram to identify the overlap degree of driver genes detected by different methods and the proposed algorithm. We have studied the degree of overlap in two aspects.

Results for comparing the overlap of detected genes by various methods for breast cancer is shown in Figure 8(a).

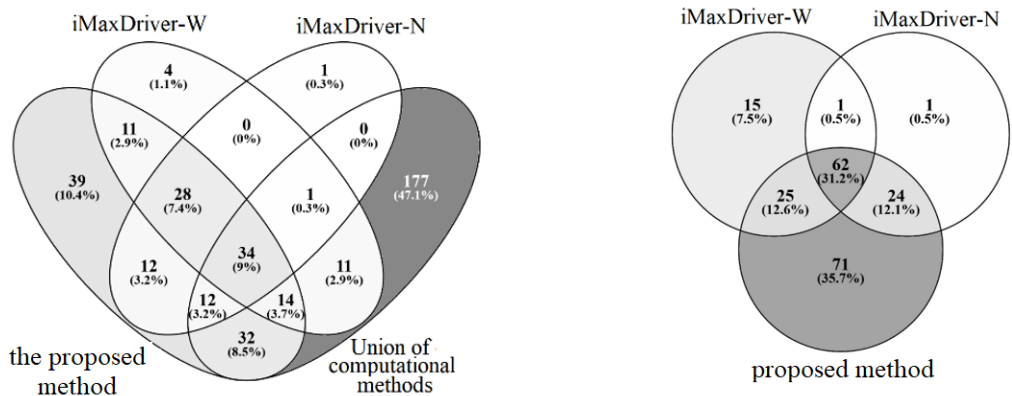

Figure 8. Overlap of detected genes in the proposed algorithm and other methods (a), Overlap of detected genes in the proposed algorithm and network-based methods (b)

In breast invasive carcinoma, the proposed algorithm was able to cover 146 driver genes that had been identified by other methods. Also, 39 new driver genes were identified that had not been identified through any of the computational or network-based methods. It was also able to identify 90 unique genes that had not been identifiable through any of the computational methods. Results indicate that network-based tools can be supplementary used alongside computational methods for the prediction of cancer driver genes. As depicted in Figure 8(b), the proposed algorithm is superior in this regard, to the two other network-based methods. The proposed method is capable of covering 111(over 86%) genes in breast invasive carcinoma, which have been identified by the two other network-based methods. In addition, 71 (over 39%) unique genes, which had not been identified by any of the two other network-based methods, were identified using the proposed algorithm, which indicates the superiority of the proposed method. In addition to the conducted evaluations, we have compared the proposed algorithm and 23 other methods regarding the percentage of predicted driver genes in various standard datasets. Figure 10 shows the related results in the CGC driver gene database for the proposed algorithm and 23 other methods. The proposed algorithm detected the highest percentage of driver genes and ranked second after iPac [8]. Figures 12 show the results for the percentage of diagnostic genes through various methods in Mut-driver standard dataset.

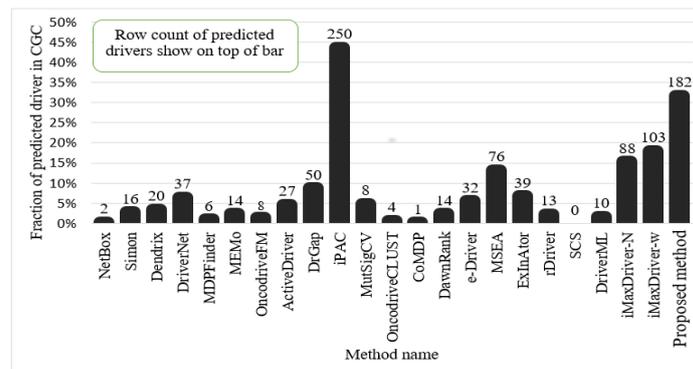

Figure 10. Fraction of predicted driver genes that are found in the Cancer Gene Census (CGC)

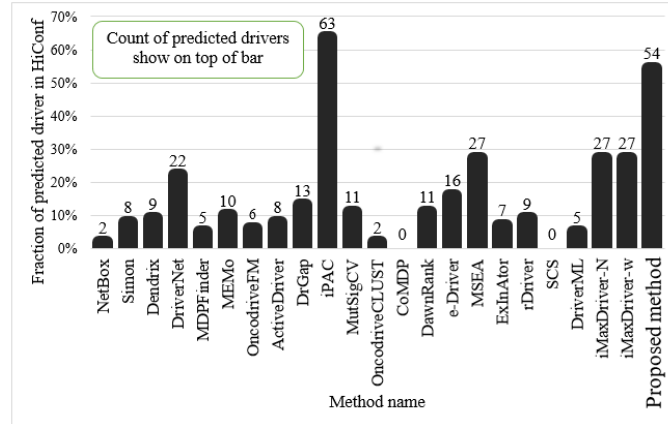

**Figure 12.** Fraction of predicted driver genes that are found in the Mut-driver dataset

Also, we have evaluated the proposed method and 23 other methods regarding the number of predicted driver genes in various standard datasets for breast cancer, the results of which are shown in Table 3. We have highlighted two highest-ranking methods for various types of cancerous tissues. As observed, the proposed algorithm is one of the two superior methods.

**Table 3.** Performance of the 23 cancer driver gene prediction methods in various datasets of breast invasive carcinoma

|  | F-measure rank | CGC rank | Mut-Driver rank |
|---|---|---|---|
| NetBox [17] | 21 | 21 | 20 |
| Simon [4] | 13 | 13 | 10 |
| Dendrix [15] | 14 | 12 | 11 |
| DriverNet [18] | 5 | 8 | 7 |
| MDPFinder [19] | 19 | 19 | 19 |
| MEMo [20] | 10 | 14 | 13 |
| OncodriveFM [5] | 18 | 18 | 18 |
| ActiveDriver [6] | 9 | 11 | 14 |
| DrGap [7] | 12 | 6 | 6 |
| iPAC [8] | 20 | 1 | 1 |
| MutSigCV [9] | 7 | 10 | 9 |
| OncodriveCLUST [10] | 6 | 20 | 21 |
| CoMDP [21] | 22 | 22 | 22 |
| DawnRank [22] | 15 | 15 | 17 |
| e-Driver [11] | 8 | 9 | 8 |
| MSEA [12] | 4 | 5 | 3 |
| ExInAtor [13] | 11 | 7 | 12 |
| rDriver [14] | 16 | 16 | 15 |
| SCS [23] | 22 | 23 | 23 |
| DriverML [16] | 17 | 17 | 16 |
| iMaxDriver-W [24] | 3 | 3 | 5 |
| iMaxDriver-N | 2 | 4 | 4 |
| Proposed method | 1 | 2 | 2 |

## 7. Conclusion

In this research, we have proposed a network-based algorithm for predicting cancer-causing genes in gene regulatory networks. The suggested approach can rank genes through structural analysis of weighted gene interactions. Therefore, the list of regulatory interactions and gene expression data for breast invasive carcinoma was downloaded, and after the required preprocessing, its network was constructed. In the next step, the nodes were weighted based on the gene expression values. Therefore, the regulatory interactions were weighted by using the network's biological and topological characteristics. Finally, by running the proposed algorithm on breast regulatory network, the list of CDGs was obtained. The majority of driver gene prediction methods rely on mutation rate and genomic data. These methods, presented in the introduction section, are categorized into three groups. A group of methods only use the concept of mutation. Another group has used network features and machine learning, alongside the concept of mutation. Other methods such as iMaxDriver-N and iMaxDriver-W have also used the network concept. Most computational methods involve a large amount of calculation and low F-measures. Even the iMaxDriver methods, which are in better shape regarding correctness criteria, involve a large amount of calculation and are very slow. Also, the weight of interactions in these methods is considered randomly. Besides, they only identify TF driver genes and are not capable of identifying driver genes of the mRNA type. However, the proposed method merely uses gene expression data and the list of regulatory interactions, without the need for mutation data. Results revealed that the proposed algorithm has a higher F-measure in comparison to other methods. The majority of methods proposed for predicting driver genes are only capable of correctly detecting a small number of them; though some have been successful in identifying a large number of driver genes, they are not of acceptable precision. For instance, iPac [8] has succeeded in identifying 4821 driver genes for breast invasive carcinoma, whereas its precision is only 6%, and its F-measure is 0.003. However, the proposed algorithm has successfully identified 858 driver genes, with a precision value of 19.5% and an F-measure of 0.234. The proposed algorithm and 23 other methods were evaluated based on five standard databases and the harmonic mean (F- measure), which considers both Recall and Precision criteria. Results showed that the proposed algorithm was one of the two superior methods in all evaluations. Also, in contrast to the other compared network-based methods, it can simultaneously identify both types of TF and mRNA driver genes. Furthermore, the proposed algorithm is capable of identifying a significant number of unique driver genes that have not been identified by other methods, showing that the proposed algorithm, as a network-based method, can be used as a supplement to advanced computational methods.

**Future work**

The Markov models can be used to analyze the behavior of genes and their signaling pathways between gene interactions in dynamic gene networks. Progressive change in target gene expression, by a specific gene in the network and through an interaction, can be modeled using first or second-order Markov chain models and utilized for predicting the probable changes in gene expression in the future.

# References

[1] L. Ding, et al., Somatic mutations affect key pathways in lung adenocarcinoma, Nature 455 (7216) (Oct. 2008) 1069–1075.
[2] L. Mularoni, R. Sabarinathan, J. Deu-Pons, A. Gonzalez-Perez, N. Lopez-Bigas, OncodriveFML: a general framework to identify coding and non-coding regions with cancer driver mutations, Genome Biol. 17 (1) (2016) 128.
[3] J. Reimand, O. Wagih, G.D. Bader, The mutational landscape of phosphorylation signaling in cancer, Sci. Rep. 3 (2013) 2651.
[4] A. Youn, R. Simon, Identifying cancer driver genes in tumor genome sequencing studies, Bioinformatics 27 (2) (2010) 175–181.
[5] A. Gonzalez-Perez, N. Lopez-Bigas, Functional impact bias reveals cancer drivers,Nucleic Acids Res. 40 (21) (2012) e169–e169.
[6] J. Reimand, O. Wagih, G.D. Bader, The mutational landscape of phosphorylation signaling in cancer, Sci. Rep. 3 (2013) 2651.
[7] Hua X., Xu H., Yang Y., Zhu J., Liu P., Lu Y.. DrGaP: A powerful tool for identifying driver genes and pathways in cancer sequencing studies. Am. J. Human Genet. 2013; 93:439–451.
[8] M.R. Aure, et al., Identifying in-trans process associated genes in breast cancer by integrated analysis of copy number and expression data, PLoS One 8 (1) (2013) e53014.
[9] M.S. Lawrence, et al., Mutational heterogeneity in cancer and the search for new cancer-associated genes, Nature 499 (7457) (2013) 214.
[10] D. Tamborero, A. Gonzalez-Perez, N. Lopez-Bigas, OncodriveCLUST: exploiting the positional clustering of somatic mutations to identify cancer genes, Bioinformatics 29 (18) (2013) 2238–2244.
[11] E. Porta-Pardo, A. Godzik, e-Driver: a novel method to identify protein regions driving cancer, Bioinformatics 30 (21) (2014) 3109–3114.
[12] D. Arneson, A. Bhattacharya, L. Shu, V.-P. Mkinen, X. Yang, Mergeomics: a web server for identifying pathological pathways, networks, and key regulators via multidimensional data integration, BMC Genomics 17 (1) (2016) 722.
[13] Lanzos A., Carlevaro-Fita J., Mularoni L., Reverter F., Palumbo E., Guigo R., Johnson R.. Discovery of Cancer Driver Long Noncoding RNAs across 1112 tumour Genomes: New candidates and distinguishing features. Sci. Rep. 2017; 7:41544.
[14] Wang Z., Ng K.S., Chen T., Kim T.B., Wang F., Shaw K., Scott K.L., Meric-Bernstam F., Mills G.B., Chen K.. Cancer driver mutation prediction through Bayesian integration of multi-omic data. PLoS one. 2018; 13:e0196939.
[15] F. Vandin, E. Upfal, B.J. Raphael, De novo discovery of mutated driver pathways in cancer, Genome Res. 22 (2) (Feb. 2012) 375–385.
[16] Kleinberg JM. Authoritative sources in a hyperlinked environment. Journal of the ACM (JACM). 1999 Sep 1; 46(5):604-32.
[17] Safar, Mostafa Akhavan, Babak Teimourpour, and Mehrdad Kargari. "GenHITS: A Network Science Approach to Driver Gene Detection in Human Regulatory Network Using Gene's Influence Evaluation." Journal of Biomedical Informatics (2020): 10366
[18] Han, H., Cho, J.W., Lee, S., Yun, A., Kim, H., Bae, D., Yang, S., Kim, C.Y., Lee, M., Kim, E. and Lee, S., 2017. TRRUST v2: an expanded reference database of human and mouse transcriptional regulatory interactions. Nucleic acids research, 46(D1), pp.D380-D386.